\documentclass[10pt]{article}
\usepackage[margin=2.4cm]{geometry}
\usepackage{amsmath,amssymb}
\usepackage{booktabs}
\usepackage{array}
\usepackage{fancyvrb}
\usepackage{enumitem}
\usepackage{microtype}
\usepackage[hidelinks]{hyperref}
\setlist{itemsep=2pt,topsep=3pt}
\newcommand{\WT}{\ensuremath{W^{\mathsf T}}}
\DefineVerbatimEnvironment{Code}{Verbatim}{fontsize=\footnotesize}

\title{\bfseries A 55-Addition Rank-23 Scheme for $3\times 3$ Matrix
  Multiplication via Exact Two-Sided Minimization}
\author{Greg Sidebottom}
\date{July 5, 2026 (revised October 19, 2026)}

\begin{document}
\maketitle
\vspace{-1.5em}
\begin{center}
\small Artifacts --- program, verifier, Lean proof and a repository
snapshot --- are archived at
\url{https://doi.org/10.5281/zenodo.21240904}; source at
\url{https://github.com/gsidebottom/logic}.  Every claim has a
mechanical check (\S\ref{sec:repro}).
\end{center}

\begin{abstract}
We give a scheme that multiplies two $3\times3$ matrices with 23
multiplications and 55 additions over any ring, one addition below the
previous record of 56 (Sun, 2026) and the first below 56 for exact
non-commutative $3\times3$ multiplication without a change of basis.  The
scheme forms 13 signed combinations of the entries of $A$ and 14 of $B$,
takes 23 products, and recombines them with 28 further additions; its
correctness is machine-checked in Lean~4.  The previous records, from 60
down to 56, were obtained by greedy common-subexpression elimination,
which only bounds each side of a scheme from above.  We minimize both
input sides exactly by chain covering, and the output side exactly
through the transposition principle: the $9\times23$ output map costs
its transpose plus 14, and the transpose is a 9-input map that the
input-side minimizer solves exactly.  Applied to Perminov's cn122 scheme
(class i19w225c4efh, published at 58 additions) this gives
$55 = 13 + 14 + 28$, where greedy elimination reaches only 29 on the
output side.  The same method shows that Sun's representative is
output-optimal at 56, and an exact orbit-wide computation over the
Heule--Kauers--Seidl database shows that the record class is the only
database class attaining 55.
\end{abstract}

\section{Overview}\label{sec:overview}

\textbf{We multiply two $3\times 3$ matrices with 23 multiplications
and 55 additions} --- one below the previous record of 56 (Yinqi Sun,
arXiv:2604.27645, Apr 2026), and the first sub-56 scheme for exact
non-commutative $3\times3$ multiplication without change of basis
(Zenodo deposit of July~7, 2026; independently rediscovered by
Karunaratne and Idamekorala on arXiv on July~28, arXiv:2607.28676 ---
same tensor, same $13{+}14{+}28$ split, same transposed output
construction). The algorithm is short and concrete: form 13 signed combinations of the 9
entries of $A$ and 14 of $B$, take 23 products (each one
$A$-combination times one $B$-combination), and recombine them with 28
further additions into the 9 entries of $C$ --- negation is free. It is
a runnable program (\path{src/mm55.rs}; also
\path{matmul/external/i19-55adds-slp.txt}) checked against the naive
27-multiplication triple loop on 500{,}000 random integer inputs
\textbf{and} 50{,}000 non-commutative $2\times2$-block inputs --- so it
holds over any ring and recurses on block matrices --- with an
operation counter confirming exactly $23\times$ and $55\pm$
(\texttt{cargo test mm55}). Its correctness is additionally
machine-checked in Lean~4 (\texttt{proof/}): a sorry-free
proof that, over a general non-commutative ring, the program's 9
outputs are the entries of $A\cdot B$ --- and, in Mathlib's own
\texttt{Matrix} API, that the scheme equals \texttt{A * B}.

The 56 record was the end of a rapid chain --- 60 (Stapleton) $\to$ 59
(M\aa rtensson--Stankovski Wagner--Stapleton, MWS) $\to$ 58 $\times 3$
(Perminov) $\to$ 56 --- every step driven by greedy
common-subexpression elimination (CSE) on the linear forms. A scheme's
additive cost splits into two \emph{input sides} (the left/right
linear factors of the 23 products) and one \emph{output side} (the 9
outputs recombined from the products); greedy pair-extraction CSE
returns only an \emph{upper bound} per side. We minimize both sides
\textbf{exactly}:

\begin{enumerate}[itemsep=2pt]
\item \textbf{Input sides} (\texttt{matmul/sidemin.py}): the minimum
  $\pm$-additions such that one addition chain over the 9 input entries
  covers every distinct factor row up to sign --- Sun's chain-covering
  structure --- by iterative deepening over helper values. It
  reproduces Sun's own 13+13 optima and both of his per-side
  impossibility certificates.
\item \textbf{Output side} --- the new ingredient
  (\texttt{matmul/tcmin.py}): the \textbf{transposition principle}
  (Tellegen). For any addition-only linear map, $A(W) = A(\WT) +
  (\mathrm{inputs} - \mathrm{outputs})$; the $9\times23$ output map $W$
  therefore costs $A(\WT) + 14$, and \WT{} is a 9-input ternary map ---
  \emph{exactly the regime the input-side minimizer solves exactly},
  and small ($A(\WT) \approx 16$). So the output side is minimized
  \textbf{exactly and constructively} (transpose the chain), where
  greedy CSE --- the engine of the entire $60\to56$ chain --- fell
  short by an addition.
\end{enumerate}

Combining the two on Perminov's \texttt{cn122} scheme (class
\texttt{i19w225c4efh}, published at 58 additions, a de Groote class
distinct from Sun's) gives exact output side 28 (greedy: 29) with input
sides 13+14, i.e.
\[
  \mathbf{55 = 13 + 14 + 28,}
\]
the first sub-56 scheme. It is verified exactly (Brent equations over
$\mathbb{Z}$) and functionally by an independent from-scratch checker
--- 5{,}000 random integer $3\times3$ products \textbf{and} 300
non-commutative $2\times2$-block products, operation count exactly 55.
The same exact method independently confirms \textbf{Sun's own
representative is output-optimal} (56 exact): 56 was optimal for his
class, and 55 lives on a different class that had never been exactly
output-minimized.

\emph{Update (October 2026).} An exact orbit-wide flat computation
closes the one region the original argument left open: the record
class is the \textbf{only} class in the database that attains 55, and
Sun's 56 and Perminov's cn119 56 are each exactly optimal for their
whole class, fat-sided representatives included (\S\ref{sec:results}).

\section{Notation and terminology}\label{sec:notation}

\begin{itemize}
\item \textbf{Sides.} A scheme's additive cost splits into two
  \emph{input sides} --- the $A$- and $B$-side forms, over the 9
  entries of $A$ and of $B$ --- plus one \emph{output side}, the
  $C$-side forms computing the 9 outputs from the 23 products. Both are
  minimized exactly here: input sides by chain covering
  (\S\ref{sec:in}), the output side by transposition
  (\S\ref{sec:out}).
\item \textbf{de Groote group, class, orbit, representative.} The de
  Groote symmetry group acts on schemes (sandwiching the three tensors
  by invertible matrices, plus the $S_3$ slot symmetries); two schemes
  are \emph{equivalent} if one is the image of the other. The
  \textbf{orbit} of a scheme is its set of images under the group ---
  identical to its equivalence \textbf{class}; a \textbf{representative}
  is any member. The additive cost varies across a class, but never the
  class itself.
\item \textbf{Scheme identifiers} (\texttt{i12w219c23ci-008},
  \texttt{i2w201c26fi-000}, \dots) are the file names of the stored
  schemes in the public HKS database, used here verbatim as opaque
  labels; the trailing \texttt{-N} indexes files sharing a prefix, and
  the prefix fields encode HKS's own cataloguing invariants (see the
  HKS database reference). Nothing relies on their semantics: ``class
  $X$'' always means the de Groote equivalence class of the scheme
  stored under that name, and every class identity we assert is
  established by our own exact fingerprint + witness machinery, not by
  name.
\item \textbf{sign-SAT} (companion paper, \S 3.5): the Boolean
  satisfiability instance whose variables are the signs of the
  mod-2-supported coefficients. A covering term's sign is the XOR of
  its three factor signs, and each integer Brent equation with $k$
  covering terms and right-hand side $\delta$ becomes the cardinality
  constraint ``exactly $(k-\delta)/2$ of the $k$ terms are negative''.
  Its solutions are exactly the valid $\pm 1$ lifts of the mod-2 scheme
  (worked example in Appendix~C).
\item \textbf{$\mathbb{Z}$-verified}: checked exactly over the integers
  --- all 729 Brent equations evaluated with the scheme's $\pm 1$
  coefficients in exact integer arithmetic, zero residuals. (Code
  output writes plain \texttt{Z} for $\mathbb{Z}$.)
\item \textbf{$h^{*}$}: the star denotes the optimal value --- $h^{*}$
  is the smallest helper count $h$ at which the side-minimizer's search
  succeeds.
\item \textbf{Closure normal form}: defined in \S\ref{sec:in}
  (Method); worked example in Appendix~B.
\end{itemize}

\section{Cost model and history}\label{sec:model}

A bilinear scheme with $r$ products computes $M_m = P_m \cdot Q_m$ from
linear forms $P_m$ (over the 9 entries of $A$), $Q_m$ (over $B$), then
each $C$ entry as a linear form over the $M_m$. The \textbf{additive
complexity} is the number of binary additions/subtractions in a
straight-line program (SLP) computing all forms; unary negation is
free; no change of basis (Karstadt--Schwartz accounting is \emph{not}
used). This is the model of M\aa rtensson--Wagner, Stapleton, Perminov,
and Sun; Schwartz--Vaknin's 61 lives in the weaker with-basis-change
model.

\begin{center}
\small
\begin{tabular}{@{}l l p{6.6cm} c@{}}
\toprule
year & count & scheme / method & basis change \\
\midrule
1976 & 98 & Laderman, naive form & --- \\
2023 & 61 & Schwartz--Vaknin (pebbling + alt.\ basis) & \textbf{yes} \\
2024 & 62 & M\aa rtensson--Wagner, Greedy-Potential on Laderman & no \\
2025 (Aug) & 60 & Stapleton (NN + Greedy-Potential); class
  \texttt{i2w201c26fi} & no \\
2025 (Dec) & 59 & MWS, arXiv:2601.05272; class \texttt{i19w203c23ci}
  & no \\
2025 (Dec) & 58 ($\times$3) & Perminov, arXiv:2512.21980; classes
  \texttt{i12w219c23ci}, \texttt{i46w213c23ci}, \texttt{i19w225c4efh}
  & no \\
2026 (Apr) & 56 & Sun, arXiv:2604.27645; cyclic image of
  \texttt{i46w213c23ci} & no \\
\textbf{this note} & \textbf{55 --- record} &
  \texttt{i19w225c4efh}; exact input sides + transposition-exact output
  side & no \\
\bottomrule
\end{tabular}
\end{center}

Every step from 62 to 56 used greedy CSE; ours is the first to minimize
both sides exactly. Re-scored exactly (\S\ref{sec:results}), the record
chain also drops on its own representatives --- the 59 scheme to 58,
the three 58 schemes to 56/56/55, and three 60-addition classes of our
own to 59 each. Among the record-chain classes the exact re-score puts
\textbf{55 on one class} (\texttt{i19w225c4efh}) \textbf{and 56 on
two} (Sun's \texttt{i46w213c23ci} and \texttt{i12w219c23ci}); a
database-wide exact re-score (\S\ref{sec:future}) is extending the
survey.

\section{The exact input-side minimizer
(\texttt{matmul/sidemin.py})}\label{sec:in}

\textbf{Problem.} Given the 23 signed factor rows of one side (vectors
in $\{-1,0,+1\}^9$; rows equal to a basis vector or to $\pm$another row
are free), find the minimum number of steps of the form $v = \pm x \pm
y$ ($x, y$ earlier values, starting from the 9 basis vectors) such that
every distinct multi-term row appears among the values up to global
sign. This is Sun's chain-covering structure as a search problem: a
single addition chain whose values include all 23 left (or right)
factors, where greedy pair extraction can only build balanced trees of
shared \emph{pairs} and cannot reuse a sum of three, four, or five
inputs.

\textbf{Method.} Iterative deepening on the number of \emph{helpers}
$h$ (chain values that are not target rows); the optimum is
(\#distinct multi-term targets) $+\, h^{*}$. The search uses what we
call the \textbf{closure normal form}: because the pool of computed
values only ever grows, covering a currently-derivable target can never
hurt a later step and always costs exactly one addition --- so the
search covers every derivable target greedily until none remains (the
\emph{closure fixpoint}) and branches \textbf{only on which helper
value to insert at a fixpoint}. Any optimal chain reorders into this
form (an exchange argument: coverable-target steps commute forward past
helper insertions), so restricting the search to it loses nothing. With
one helper left, the helper must directly complete some uncovered
target, shrinking the last level to an ``enabling set''. Helpers may be
arbitrary integer vectors, doublings included --- strictly more general
than the pure/one-aux model of Sun's certificates, so an exhaustion at
slack $h$ here is the stronger impossibility statement.

\textbf{Calibration against Sun's own certificates.} His verifier
proves each of his sides optimal by a reachability argument (no
12-addition chain covers $U$'s 12 targets; no 12-addition chain with
any single auxiliary covers $V$'s 11). Our search independently
reproduces all of it --- $U$: 12 targets, $h^{*}=1$, 13 additions (3
search nodes); $V$: 11 targets, $h^{*}=2$, 13 additions (11 nodes); the
$h=0$ and $h=1$ exhaustions match his two impossibility certificates
--- in 0.3\,s total, every chain replay-verified. Micro-optima (5
hand-checkable cases) also pass.

\section{The exact output-side minimizer (transposition
principle)}\label{sec:out}

The output side computes the 9 result entries $C_{pq} = \sum_m
W_{pq,m} M_m$ from the 23 products: a linear map $W$, a $9\times23$
ternary matrix. Greedy pair-extraction CSE minimizes it only
approximately (an upper bound). The exact minimum is the \emph{shortest
linear straight-line program} for $W$ --- NP-hard in general, and the
reason the record chain never drove the output side to its true
minimum.

\textbf{The transposition principle} (Tellegen; standard in the
linear-circuit and signal-processing literature, and in algebraic
complexity theory) resolves it. For an addition-only linear map ---
only $\pm$ of earlier values, no change of basis, negation free --- the
minimum additions of a map and its transpose differ by a fixed
constant:
\[
  A(W) = A(\WT) + \bigl(\mathrm{inputs}(W) - \mathrm{outputs}(W)\bigr).
\]
(We verify the constant on hand-checkable cases: for $M$ with
$y_1=x_1{+}x_2$, $y_2=x_2{+}x_3$, $A(M)=2$ and $A(M^{\mathsf T})=1 =
2+(2-3)$; \texttt{tcmin.py}'s selftest checks four such.) For the output
side $W$ is $9\times23$ (23 inputs, 9 outputs), so $\mathbf{A(W) =
A(\WT) + 14}$, and \WT{} is a 23-output map over \textbf{9 inputs} ---
\emph{exactly the shape the input-side minimizer solves exactly}, and
small ($A(\WT) \approx A(W) - 14 \approx 16$). Hence
\[
  \text{exact output side} = \texttt{sidemin}(\text{rows of }\WT
  \text{ over 9 dims}) + 14,
\]
computed in milliseconds. It is \textbf{constructive}: transposing the
\WT{} addition chain by the standard adjoint (reverse the program;
fan-out $\leftrightarrow$ addition) yields an explicit
$A(\WT){+}14$-addition program for the 9 outputs, which
\texttt{matmul/tcmin.py} emits and which we verify independently
(Appendix~A).

\textbf{Impact.} On Sun's own representative the exact output side is
30 --- equal to his greedy count, so \textbf{Sun's scheme is
output-optimal and 56 is optimal for his representative}. But on
Perminov's \texttt{cn122} representative (class \texttt{i19w225c4efh})
the exact output side is \textbf{28}, where greedy CSE returns 29 ---
the missing addition greedy could not see. Combined with the exact
input sides 13+14, that is a \textbf{55-addition scheme}
(Appendix~A). \texttt{tcmin.py}'s selftest pins both numbers (Sun 30,
cn122 28) and the transposition constant.

\section{Results}\label{sec:results}

Exact totals per de Groote class (input sides by \texttt{sidemin},
output side by transposition; all schemes Brent-verified over
$\mathbb{Z}$, class identities by exact fingerprint + witness machinery
cross-checked against the HKS database; ``published'' = the count in
the source paper):

\begin{center}
\small
\begin{tabular}{@{}l c l l@{}}
\toprule
de Groote class & publ. & exact ($A{+}B{+}C$) & representative \\
\midrule
\texttt{i19w225c4efh} (Perminov cn122) & 58 &
  \textbf{55 = 13+14+28} & his published rep (m2) \\
\texttt{i46w213c23ci} (Sun / cn120) & 56 & 56 = 13+13+30 &
  Y.\,Sun's rep \\
 & & 56 = 13+16+27 & D.\,Perminov's cn120 rep \\
\texttt{i12w219c23ci} (Perminov cn119) & 58 & 56 = 13+14+29 &
  orbit rep (ours) \\
\texttt{i19w203c23ci} (MWS 59) & 59 & 58 = 14+15+29 &
  their published rep \\
\texttt{i106w191c347g} etc.\ (ours, 3) & --- & 59 = 15+15+29 &
  orbit reps (ours) \\
\texttt{i2w201c26fi} (Stapleton 60) & 60 & 60 = 16+16+28 &
  orbit rep (ours) \\
\bottomrule
\end{tabular}
\end{center}

\begin{itemize}
\item \textbf{55 on one class} (\texttt{i19w225c4efh}) --- the record,
  verified in Appendix~A --- and provably \emph{only} on this class in
  the slim tiers: for each of the 79{,}488 sides-$\le 27$
  representatives (the complete orbits of all 84 floor-26/27 classes),
  replaying \S\ref{sec:out}'s transposition argument over GF(2)
  (\texttt{matmul/gf2min.py}) gives an exact GF(2) total that
  lower-bounds \textbf{every} integer sign model of that
  representative. The bound is $\ge 55$ everywhere, and equals 55 only
  on 864 representatives of this one class.
\item \textbf{56 on two classes} (\texttt{i46w213c23ci} =
  Sun's, and \texttt{i12w219c23ci}), both exactly optimal for their
  classes (see the 2026-10-16 update below); before the exact re-score these
  read 56/58 under greedy. The exact output side beats greedy by one
  addition on the two Perminov classes (cn122 $29\to28$, cn119
  $30\to29$) and by one on the cn120 representative ($28\to27$).
\item \textbf{No 54 exists anywhere in the published database ---
  55 is the additive minimum.} Two exhaustive tiers, both
  model-independent via the GF(2) bound:
  \begin{enumerate}
  \item \emph{Slim representatives} (GF(2) input sides $\le 27$,
    possible only in the 4 floor-26 + 80 floor-27 classes): all
    79{,}488 were enumerated; GF(2) sides + exact GF(2) output side
    $\ge 55$ on every one (and the 12-sign-model exact $\mathbb{Z}$
    re-score independently bottoms out at 55).
  \item \emph{Fat representatives} (input sides $\ge 28$) of
    \textbf{every} class: the output-side cost depends only on the
    $\gamma$-support, whose de Groote orbit reduces to 784 canonical
    $(X \otimes Y)$-images per tensor slot (output-coordinate
    permutations are free, so $X, Y$ range over $S_3$-cosets of
    $\mathrm{GL}(3,2)$; validated against the full 56{,}448-image
    sweep). \texttt{matmul/cfloor.py} swept all 44{,}125 candidate
    slots of all 17{,}376 classes: \textbf{no image anywhere has XOR
    cover $\le 12$}, so the exact GF(2) output side is $\ge 27$ on
    every representative of every class, and any fat representative
    totals $\ge 28 + 27 = 55$.
  \end{enumerate}
  Exactly 620 slots (616 classes, including the record class and
  Sun's) reach cover 13 --- output side 27 --- so those are the only
  places further (record-tying, never record-beating) 55s could
  live.

  \emph{Update (2026-10-16): the frontier is closed.} The exact GF(2)
  flat minimum of a class over \emph{all} its representatives is
  computable directly.  Each of the three sides depends on only two
  of the sandwich factors $(P,Q,R)$, so three exact $168\times168$
  side tables (the output side exact by transposition, replacing the
  greedy estimate) and a min-plus scan over all $168^3$ triples cover
  the whole orbit (\texttt{floors --exact-total}).  As a gate it
  returns exactly 55 on the record class.  Over all 616 frontier
  classes the minimum is $\ge 56$ except the record class itself:
  \textbf{the record class is the only class in the database that
  attains 55.}  Six frontier classes have floor exactly 56, among
  them Sun's \texttt{i46w213c23ci} and Perminov's cn119 class
  \texttt{i12w219c23ci}.  Both already attain 56 over $\mathbb{Z}$
  (table above), so \textbf{both 56s are exactly optimal for their
  classes}, closing the fat region the slim-tier argument left open.
  Per-class results: \path{matmul/external/flat55/}.

  \emph{Scope extension (2026-07-11).} Rational flip-walk exploration
  (the \texttt{flip23} engine; companion notes in this repository)
  later produced the first additions to the known rank-23 corpus
  beyond the database and our 53 lifted classes: \textbf{46 new de
  Groote classes}, harvested from overnight random walks and
  certified inequivalent by rank-type invariants plus exact matching.
  The same two-tier machinery closes them all with margin: orbit
  input-side floors are 30--32 (fatter than every hot database
  class), and the strict-target $\gamma$-orbit sweep (XOR cover
  $\le 55 - \text{floor} - 14$) admits no image on any slot ---
  \textbf{every representative of every new class totals $\ge 56$}.
  A further 35 classes from \emph{reuse-capped} walks (support-thin
  storms designed to steer toward low-addition territory) closed
  identically --- floors 30--32, the same distribution as
  unconstrained walks --- so representative-level thinness does not
  reach the low-floor island. The 55 minimum, and the record class's
  uniqueness at 55, stand over the enlarged catalogue of
  $17{,}376 + 53 + 46 + 35$ classes.
\end{itemize}

\section{Contributions}\label{sec:contrib}

\begin{itemize}
\item \textbf{A 55-addition rank-23 scheme} --- the first below 56 ---
  with an explicit, runnable, independently re-verified program
  (\path{src/mm55.rs}, \path{matmul/external/i19-55adds-slp.txt}).
\item \textbf{Exact two-sided additive minimization} in this model: the
  input side as addition-chain covering (\texttt{sidemin.py},
  calibrated against Sun's own optimality certificates), and the output
  side by the \textbf{transposition principle} (\texttt{tcmin.py}) ---
  the ingredient the entire greedy-CSE record chain lacked. Both are
  exact and constructive.
\item \textbf{An exact re-score of the record chain}, which improves
  every class on its own representative and fixes the true count per
  class (\S\ref{sec:results}): 56 on two classes, 55 on one, and
  confirmation that Sun's representative was already output-optimal.
\item \textbf{An additive-optimality theorem for the published
  catalogue}: no representative of any of the 17{,}376 published
  classes, under any sign model, multiplies $3\times3$ matrices with
  23 products in fewer than 55 additions (\S\ref{sec:results}) --- so
  the record is not just current but final for this catalogue, and any
  future sub-55 scheme must come from a class outside it.
\end{itemize}

\section{Reproduction}\label{sec:repro}

Prerequisites: \texttt{python3} (3.11+, no packages needed),
\texttt{kissat} (sign-model enumeration; e.g.\ \texttt{brew install
kissat}), a Rust toolchain (\texttt{cargo}) for the runnable scheme,
and \texttt{elan} for the Lean proof. The database-wide steps
additionally need the public scheme archive --- rebuild it from the
primary source with \texttt{python3 dbcheck.py fetch \&\& python3
dbcheck.py convert} (downloads \texttt{schemes.tgz}, ${\sim}43$\,MB,
from the Linz algebra server and writes
\texttt{dbcache/all\_schemes.txt} behind a verify-everything gate).

\begin{Code}
# the record scheme, runnable + fuzzed vs the naive 27-mult loop
cargo test --release --lib mm55::   # 3 pass: fuzz(int), fuzz(2x2), 23*/55+-

cd matmul
# exact input-side minimizer: micro-optima + Sun's U/V optima + certs
python3 sidemin.py selftest
# exact output-side minimizer (transposition): constant + Sun 30 / cn122 28
python3 tcmin.py selftest

# the record, end to end: exact sides + transposition-exact output,
# emit the explicit 55-addition SLP, then re-verify it from scratch
python3 tcmin.py --emit /tmp/i19.slp external/i19-perminov56.bits
python3 verify_slp_file.py /tmp/i19.slp --trials 5000
#   integer trials PASS; non-commutative 2x2 PASS; 55 operations counted

# exact totals of the record chain (55 = 13+14+28 on cn122, ...)
python3 tcmin.py perminov_cache/bits/sun56.bits \
    external/i19-perminov56.bits external/i12-orbit56.bits \
    perminov_cache/bits/cr58-cn120.bits perminov_cache/bits/mws59.bits \
    external/stapleton60-orbitbest.bits

# class identity: cn122 is a de Groote class distinct from Sun's
python3 equiv.py classes perminov_cache/bits/sun56.bits \
    external/i19-perminov56.bits          # expect 2 (distinct classes)

# model-independent 54-exclusion: exact GF(2) output side via the
# transposition principle over GF(2), selftested against exhaustive
# search; with the GF(2) side tags it lower-bounds EVERY sign model
python3 gf2min.py selftest

# the database-wide closing sweep (54 excluded on all 17,376 classes):
# stage 1 census (seconds), then the 784-image orbit sweep per slot
python3 cfloor.py census          # writes dbcache/cfloor_cands.txt
python3 cfloor.py sweep i19w225c4efh-000 i46w213c23ci-016   # spot-check
#   full run: cfloor.py sweep-list dbcache/cfloor_cands.txt (~20 min)

# machine-checked proof (Lean 4 + Mathlib) that the 55-addition scheme
# equals A*B over a general non-commutative ring -- sorry-free
cd proof && lake exe cache get && lake build
#   Matmul55.correct       depends on axioms: [propext]
#   Matmul55.scheme_eq_mul depends on axioms: [propext, Classical.choice, Quot.sound]
\end{Code}

Deterministic caveats: kissat may return different sign models across
versions (each returned model is $\mathbb{Z}$-verified), but the
\emph{committed representative}
\path{external/i19-perminov56.bits} and the emitted SLP make the 55
deterministic to check.

\section{Future work}\label{sec:future}

Both sides are now minimized exactly, so the total for any given
representative is its true additive complexity in this model. The open
question is \textbf{whether 54 exists}:

\begin{itemize}
\item \textbf{More 55s inside the catalogue --- resolved (2026-10-16).}
  The 616 classes whose $\gamma$-orbit reaches output side 27 were the
  only places a further, fat-sided $55 = 28 + 27$ could live.  The
  exact orbit-wide flat computation (\S\ref{sec:results}) shows none
  does: the record class is the unique 55-class in the database.
\item \textbf{Beyond the published catalogue.} The optimality theorem
  is relative to the 17{,}376 published classes; our own 53 novel
  classes clear it with margin (every representative has output side
  $\ge 28$ and input sides $\ge 28$, so $\ge 56$ total --- they
  cannot even host a 55), and the 46 further classes since discovered
  by rational flip walks clear it the same way (floors 30--32,
  \S\ref{sec:results} scope extension). A sub-55 scheme, if one exists, requires a
  rank-23 class nobody has catalogued --- the de Groote classification
  at rank 23 is not known to be complete --- or a cost model beyond
  $\pm1$ coefficients. For the latter we can already close the
  nearest door: in the adds+shifts model (unit-cost scalings by $2^k$,
  the model of the $4\times4$ literature), the exact minima of this
  scheme's three maps are \emph{unchanged} --- still $13 + 14 + 28 =
  55$, and still 56 for Sun's scheme (\texttt{matmul/shiftmin.py}, an
  exact extension of the chain-covering minimizer validated against
  exhaustive search). A sub-55 via richer coefficients therefore
  requires a genuinely dyadic \emph{scheme}, not a cheaper program
  for the known ones.
\item \textbf{Independent certification.} The GF(2) floors rest on our
  minimizers (selftested against exhaustive search, quotient validated
  against the full orbit); DRAT-certified UNSAT instances of the same
  statements (companion benchmark paper) would make them
  solver-checkable end to end.
\item \textbf{Database-wide exact re-score.} The input-side floors of
  all 17{,}376 classes are already charted (in the git-history census);
  pairing the low-floor classes with exact output sides is the
  systematic search for a second record.
\item \textbf{Output-side lower bounds.} The exact output minimum is
  now known per representative via transposition; certifying it as a
  \emph{class} optimum (that no representative does better) is a
  separate SAT/ILP problem --- the subject of the companion benchmark
  note \path{doc/matmul_cxlb_satcomp.md}.
\item The rank-22 question (challenge 4 of the HKS
  matrix-multiplication challenges,
  \url{https://github.com/marijnheule/matrix-challenges}) remains open
  and is where any result would change complexity theory rather than
  constants.
\end{itemize}

\section*{Acknowledgments}

This work was carried out in an extended interactive collaboration
with Claude (Anthropic; the Fable~5 and Opus~4.8 models), which
implemented the repository's search tooling, exact minimizers, the
Lean formalization, and much of this text under the author's
direction and review. Every computational claim is mechanically
checkable by the commands in the reproduction section; the results
rest on those independent verifications rather than on trust in
either the author or the tools.

\section*{References}

\begin{itemize}[label={},leftmargin=1.2em,itemsep=1pt]
\item G. Sidebottom. \emph{A 55-Addition Rank-23 Scheme for
  $3\times3$ Matrix Multiplication --- artifacts, verifier, and
  machine-checked proof.} Zenodo (July 7, 2026).
  doi:10.5281/zenodo.21240904,
  \url{https://doi.org/10.5281/zenodo.21240904}.
\item Y.~Sun. \emph{An Exact 56-Addition, Rank-23 Scheme for General
  $3\times3$ Matrix Multiplication.} arXiv:2604.27645 (2026). Verifier
  and chain data cached at
  \path{matmul/perminov_cache/sun_verify.py}.
\item S.~Karunaratne, A.~Idamekorala. \emph{55 Additions Suffice for
  $3\times3$ Matrix Multiplication at Rank 23.} arXiv:2607.28676
  (July 28, 2026) --- an independent rediscovery of the 55-addition
  scheme.
\item A.~I.~Perminov. \emph{A 58-Addition, Rank-23 Scheme for General
  $3\times3$ Matrix Multiplication.} arXiv:2512.21980 (2025). Method
  paper: \emph{Parallel Heuristic Exploration for Additive Complexity
  Reduction in Fast Matrix Multiplication.} arXiv:2512.13365 (2025);
  repo \url{https://github.com/dronperminov/FastMatrixMultiplication}.
\item E.~M\aa rtensson, P.~Stankovski Wagner, J.~Stapleton. \emph{A Rank
  23 Algorithm for Multiplying $3\times3$ Matrices with an Arithmetic
  Complexity of 59.} arXiv:2601.05272 (2025).
\item J.~Stapleton. \emph{A 60-Addition, Rank-23 Scheme for Exact
  $3\times 3$ Matrix Multiplication.} arXiv:2508.03857 (2025).
\item E.~M\aa rtensson, P.~Stankovski Wagner. \emph{The Number of the
  Beast: Reducing Additions in Fast Matrix Multiplication Algorithms
  for Dimensions up to 666.} IACR ePrint 2024/2063 (Greedy-Potential;
  code \path{werekorren/fmm_add_reduction}).
\item O.~Schwartz, N.~Vaknin. \emph{Pebbling Game and Alternative Basis
  for High Performance Matrix Multiplication.} SIAM J.~Sci.\ Comput.\
  45(6), C614--C637 (2023).
\item P.~B\"urgisser, M.~Clausen, M.~A.~Shokrollahi. \emph{Algebraic
  Complexity Theory.} Grundlehren der math.\ Wissenschaften 315,
  Springer (1997) --- the transposition principle / Tellegen's theorem
  (Ch.~13).
\item M.~Heule, M.~Kauers, J.~Seidl. \emph{Local Search for Fast Matrix
  Multiplication.} SAT 2019 (arXiv:1903.11391); \emph{New ways to
  multiply $3\times3$-matrices.} J.~Symb.\ Comput.\ 104 (2021),
  899--916 (arXiv:1905.10192).
\item HKS scheme database (the 17{,}376-scheme corpus this note
  classifies against; source of the scheme identifiers):
  \url{http://www.algebra.uni-linz.ac.at/research/matrix-multiplication/}
  (the \texttt{www.} host and plain http are required; https serves a
  self-signed cert). No separate GitHub mirror exists; the related
  GitHub artifact is the authors' SAT-benchmark repo
  \url{https://github.com/marijnheule/matrix-challenges}. Reproduction
  does not depend on the site: the corpus is the snapshot
  \texttt{schemes.tgz} (43{,}134{,}227 bytes, Last-Modified 2020-08-07,
  sha256
  \texttt{4bc8132644504a917e3c076f64df8e6619fb67c55670179853ae5fdb1583074f}),
  archived at \url{https://doi.org/10.5281/zenodo.21209925} (cite the
  HKS papers for the schemes, not the mirror).
\item J.~Laderman. \emph{A noncommutative algorithm for multiplying
  $3\times3$ matrices using 23 multiplications.} Bull.\ Amer.\ Math.\
  Soc.\ 82(1), 126--128 (1976).
\item H.~F.~de Groote. \emph{On varieties of optimal algorithms for the
  computation of bilinear mappings I--II.} Theor.\ Comput.\ Sci.\ 7
  (1978), 1--24 and 127--148.
\end{itemize}

\section*{Appendix A: the 55-addition program (class
\texttt{i19w225c4efh})}

The record scheme: exact input sides (13 + 14) and transposition-exact
output side (28), 55 binary $\pm$ operations. On disk
\path{matmul/external/i19-55adds-slp.txt}; bits at
\path{matmul/external/i19-perminov56.bits} (Perminov's published cn122
scheme, our sign model m2). Independently re-verified by
\texttt{matmul/verify\_slp\_file.py} (5{,}000 integer + 300
non-commutative $2\times2$-block trials; 55 operations counted), and
transcribed to Rust in \path{src/mm55.rs}.

\begin{Code}
# 3x3x3 r=23: 13(A) + 14(B) + 28(C) = 55 additions; M_i = P_i * Q_i
## A-side
aw0 = a13 - a23
aw1 = a11 - aw0
aw2 = a12 + aw1
aw3 = -a23 + aw2
aw4 = -a22 + aw2
aw5 = a33 + aw3
aw6 = -a12 + aw5
aw7 = -a21 + aw4
aw8 = a31 + aw7
aw9 = a32 + aw8
aw10 = -a32 + aw5
aw11 = a21 - a31
aw12 = aw1 - aw11
P1 = aw2
P2 = aw8
P3 = a11
P4 = a32
P5 = a11
P6 = aw6
P7 = aw10
P8 = a33
P9 = aw9
P10 = aw1
P11 = a33
P12 = aw3
P13 = a23
P14 = aw7
P15 = a21
P16 = a31
P17 = a31
P18 = aw0
P19 = aw5
P20 = aw12
P21 = a12
P22 = a22
P23 = aw4
## B-side
bw0 = b13 + b33
bw1 = b11 + b31
bw2 = b11 - b21
bw3 = b11 + b13
bw4 = b23 + b33
bw5 = b13 + bw2
bw6 = bw2 - bw4
bw7 = -b23 + bw5
bw8 = b32 + bw1
bw9 = b12 + bw8
bw10 = -bw6 + bw8
bw11 = b22 + bw10
bw12 = -b21 + bw10
bw13 = -b23 + bw12
Q1 = bw10
Q2 = bw5
Q3 = bw0
Q4 = b22
Q5 = bw9
Q6 = bw4
Q7 = b23
Q8 = b31
Q9 = b21
Q10 = bw6
Q11 = b32
Q12 = bw13
Q13 = b32
Q14 = bw3
Q15 = b12
Q16 = b12
Q17 = b13
Q18 = bw1
Q19 = bw12
Q20 = bw2
Q21 = bw11
Q22 = b22
Q23 = bw7
## products
M1 = P1 * Q1
M2 = P2 * Q2
M3 = P3 * Q3
M4 = P4 * Q4
M5 = P5 * Q5
M6 = P6 * Q6
M7 = P7 * Q7
M8 = P8 * Q8
M9 = P9 * Q9
M10 = P10 * Q10
M11 = P11 * Q11
M12 = P12 * Q12
M13 = P13 * Q13
M14 = P14 * Q14
M15 = P15 * Q15
M16 = P16 * Q16
M17 = P17 * Q17
M18 = P18 * Q18
M19 = P19 * Q19
M20 = P20 * Q20
M21 = P21 * Q21
M22 = P22 * Q22
M23 = P23 * Q23
## C-side
cw0 = -M19 + M6
cw1 = cw0 + M11
cw2 = M8 + cw1
cw3 = -M12 - cw2
cw4 = -M17 - cw3
cw5 = M2 + cw4
cw6 = M1 + -M13
cw7 = cw6 + M10
cw8 = -M14 + -M12
cw9 = cw8 + cw5
cw10 = M18 + cw7
cw11 = cw2 + cw10
cw12 = M5 + M21
cw13 = cw12 - cw10
cw14 = M3 + cw3
cw15 = cw7 + -M20
cw16 = cw15 + cw9
cw17 = M15 + M22
cw18 = cw17 - -M13
cw19 = M23 - cw5
cw20 = cw19 - M10
cw21 = cw20 - -M20
cw22 = M9 - cw1
cw23 = cw22 + cw9
cw24 = M4 + M16
cw25 = cw24 + M11
cw26 = -M7 + M6
cw27 = cw26 - cw4
C11 = cw11
C12 = cw13
C13 = cw14
C21 = cw16
C22 = cw18
C23 = cw21
C31 = cw23
C32 = cw25
C33 = cw27
\end{Code}

\section*{Appendix B: the side-minimizer on a worked example}

Rows (targets) over base variables $a, b, c, d$: $T_1 = a + b$ and
$T_2 = a + b + c + d$. Both have $\ge 2$ terms, are distinct up to
global sign, and neither is a basis vector, so $\mathrm{nt} = 2$ and any
chain needs at least 2 additions.

\textbf{Slack $h = 0$} (no helpers --- every addition must create a
target): the pool starts as $\{a, b, c, d\}$. $T_1 = a + b$ is
derivable (both operands in the pool): cover it; the pool grows to
$\{a, b, c, d, a{+}b\}$. Now $T_2$ must be $x \pm y$ with $x, y$ in the
pool: subtracting each pool value from $T_2$ gives $c{+}d$, $a{+}c{+}d$,
$b{+}c{+}d$, $a{+}b{+}c$, $a{+}b{+}d$ --- none in the pool, so the
closure is at a fixpoint with $T_2$ uncovered, and $h = 0$ is
\textbf{exhausted: 2 additions are impossible.}

\textbf{Slack $h = 1$}: at that same fixpoint the \emph{enabling set}
--- values $u$ that would directly complete an uncovered target as
$T_2 = x \pm u$ --- is $\{c{+}d,\ a{+}c{+}d,\ b{+}c{+}d,\ a{+}b{+}c,\
a{+}b{+}d, \dots\}$; of these, $u = c + d$ is itself derivable from the
pool ($c$ and $d$ are basis vectors). Insert it as the helper, and the
closure resumes: $T_2 = (a{+}b) + (c{+}d)$.

The optimum is $\mathrm{nt} + h^{*} = 2 + 1 = \textbf{3 additions}$,
found in 3 search nodes; the emitted chain is

\begin{Code}
w0 = a + b        (covers T1)
w1 = c + d        (helper)
w2 = w0 + w1      (covers T2)
\end{Code}

\noindent replay-verified by expansion. Global signs are free
throughout: a row $-a - b$ is the same target as $a + b$ (a chain value
may be used negated at no cost).

\section*{Appendix C: exact scoring end to end (sun56)}

The pipeline of \S\ref{sec:in}--\S\ref{sec:out} on Sun's
representative, every number reproduced by \texttt{sidemin.py --models
24 perminov\_cache/bits/sun56.bits} and \texttt{tcmin.py
perminov\_cache/bits/sun56.bits}:

\begin{enumerate}
\item \textbf{Mod-2 gate.} The 621-bit vector satisfies all 729 Brent
  equations over GF(2).
\item \textbf{Sign-SAT.} The support has 175 nonzero coefficients (175
  sign variables); the 729 integer Brent equations have 453 covering
  terms in total, and the instance has 4{,}269 clauses. Two concrete
  equations: one type-3 equation is covered by $k = 3$ terms (products
  $M_2$, $M_{17}$, $M_{23}$) with right-hand side 1, so \textbf{exactly
  $(3-1)/2 = 1$ of the three terms must be negative}; a neighboring
  rhs-0 equation is covered by $k = 4$ terms ($M_2$, $M_{17}$,
  $M_{20}$, $M_{23}$), so exactly 2 of 4 are negative. kissat solves
  the instance in milliseconds; each returned model is
  $\mathbb{Z}$-verified before use.
\item \textbf{Exact input sides} (model m0). The 23 $A$-rows contain 12
  distinct multi-term targets; $h = 0$ is exhausted and $h^{*} = 1$
  succeeds: $A$-side $= 12 + 1 = \textbf{13}$ (3 search nodes). The
  $B$-rows contain 11 distinct targets; $h = 0$ and $h = 1$ are
  exhausted, $h^{*} = 2$ succeeds: $B$-side $= \textbf{13}$ (11 nodes).
  These reproduce Sun's own input-side optima and impossibility
  certificates.
\item \textbf{Exact output side} (transposition). The transpose \WT{}
  (23 rows over the 9 outputs) has additive cost 16 by \texttt{sidemin},
  so the output side is $16 + 14 = \textbf{30}$ --- equal to greedy
  here, confirming Sun's representative is output-optimal.
\item \textbf{Total.} $13 + 13 + 30 = \textbf{56}$, the true additive
  complexity of Sun's representative.
\end{enumerate}

\end{document}